\documentclass[twocolumn,aps,prd,10pt]{revtex4-1}

\usepackage{natbib}
\usepackage{bm,dcolumn,amssymb,amsmath}
\usepackage{graphicx}

\bibpunct{(}{)}{;}{a}{}{,}
\DeclareMathOperator{\erf}{erf}
\newcolumntype{d}{D{.}{.}{-1}}    %Align on decimal point

\newcommand{\element}[2]{\ensuremath{^{#1}\textrm{#2}}}
\newcommand{\ion}[2]{\textrm{#1}\,\textsc{#2}}

\def\ap{Appl. Phys.}
\def\apj{Astrophys. J.}
\def\apjss{Astrophys. J. Suppl. Ser.}
\def\lnp{Lect. Notes Phys.}
\def\mnras{Mon. Not. R. Astron. Soc.}
\def\pra{Phys. Rev. A}
\def\prl{Phys. Rev. Lett.}
\def\rmp{Rev. Mod. Phys.}

\begin{document}

\title{A reanalysis of quasar absorption spectra results suggesting a spatial gradient in values of the fine-structure constant}

\author{J. C. Berengut}
\author{E. M. Kava}
\author{V. V. Flambaum}
\affiliation{School of Physics, University of New South Wales, Sydney, NSW 2052, Australia}

\date{22 May 2012}

\begin{abstract}
We statistically analyse a recent sample of data points measuring the fine-structure constant $\alpha$ (relative to the terrestrial value) in quasar absorption systems. Using different statistical techniques, we find general agreement with previous authors that a dipole model is a well-justified fit to the data. We determine the significance of the dipole fit relative to that of a simple monopole fit, discuss the consistency of the interpretation, and test alternate models for potential variation of $\alpha$ against the data. Using a simple analysis we find that the monopole term (the constant offset in $\Delta\alpha/\alpha$) may be caused by non-terrestrial magnesium isotope abundances in the absorbers. Finally we test the domain-wall model against the data.
\end{abstract}

\maketitle

\section{Introduction}

Searching for possible variations in the fundamental constants of nature allows us to test whether the laws of physics vary over space and time. There are many complementary probes that, taken together, span much of the Universe's history~\citep[see, e.g., the review][]{uzan03rmp}. Optical quasar absorption spectra provide a probe of variations in the fine-structure constant, $\alpha=e^2/\hbar c$, along a past-light cone centred on present-day telescopes. Numerous works employed comparison of alkali-doublets in space and in the laboratory to place limits on $\alpha$-variation; these are easily interpreted since the energy separations are simply proportional to $\alpha^2$.

\citet{dzuba99pra,dzuba99prl} developed a different approach: the many-multiplet method, which exploits larger sensitivities to $\alpha$-variation available in heavier ions and in transitions other than alkali-doublets. Moreover, the method allows for better control of systematics because there exist some negative shifters (with energy intervals that reduce with increasing $\alpha$) as well as positive shifters and ``anchor lines'' (which have very weak sensitivity to $\alpha$-variation). In some cases both positive and negative shifters may be observed in the same ion. Additionally, the many-multiplet method sees a statistical gain because many more observed transitions may be used in the analysis.

During the last several years many absorption systems, observed using the Keck telescope in Hawaii~\citep{webb99prl,murphy01mnrasA,murphy03mnras,murphy04lnp} and the Very Large Telescope in Chile~\citep{webb11prl,king12mnras}, have been analysed. Taken together, these provide measurements of $\Delta\alpha/\alpha$ in $\sim300$ absorption systems covering most of the sky. Here $\Delta\alpha/\alpha=(\alpha(\vec{r}) - \alpha_0)/\alpha_0$ is the relative variation in $\alpha$ at a particular position $\vec{r}$ in the Universe where the absorption occurs. In~\citet{webb11prl}, the combined data sample is interpreted as providing evidence for variation in $\alpha$ throughout the Universe with an angular-dependence. This ``dipole'' model is found to be preferred to a monopole (constant offset) model of the variation at the $4.1\sigma$ level.

The values of $\Delta\alpha/\alpha$ for each absorber~\citep[presented in][]{murphy04lnp,webb11prl} are modelled in \citet{webb11prl}\citep[see also][]{king12mnras} using several variants of a formula which can be written as
\begin{equation}
\label{eq:dipolemonopole}
\frac{\Delta\alpha}{\alpha} = A + B(z) \cos\theta
\end{equation}
where $\theta$ is the angle between the direction of the measurement and the axis of the dipole, $A$ is a constant (a ``monopole'' term) and $B$ is the magnitude of the dipole term. The model variation comes from the form of the factor $B$, which in~\citet{webb11prl} was variously
\begin{itemize}
\item $B(z) = B_0$, i.e. Eq.~(\ref{eq:dipolemonopole}) represents pure angular-dependence without any distance-dependence;
\item $B(z) = B_0\,r(z)$, where $r=ct$ is the lookback time in giga-lightyears (Glyr);
\item $B(z) = B_0\,z^\beta$, here $z$ is the redshift and $\beta$ is another fitting parameter.
\end{itemize}
The model of spatial variation that does not include a distance dependence is difficult to understand theoretically, while fitting $z^\beta$ simply adds a poorly-constrained parameter to the data. Therefore in this paper we test the dipole interpretation $B(z)\sim r$. In this form~(\ref{eq:dipolemonopole}) represents a gradient in the value $\alpha$ throughout the Universe, and $r\cos\theta$ is the distance to a quasar absorption system along that gradient.

Of course, $r$ itself is model dependent at large redshifts. In this work we use the standard $\Lambda_\textrm{CDM}$ cosmology parametrized by WMAP5~\citep{hinshaw09apjss} to determine $r$ from the redshift $z$:
\begin{equation}
\label{eq:lookback_time}
r(z) = \frac{c}{H_0} \int_{1/(1+z)}^1 
   \frac{1}{\sqrt{\Omega_m a^{-3}+\Omega_\Lambda}}\frac{da}{a} \,.
\end{equation}
Again, this reflects the fact that we are taking measurements along a past light cone centred on present-day Earth. An alternative approach is to use the comoving distance rather than lookback time:
\begin{equation}
\label{eq:comoving_distance}
d(z) = \frac{c}{H_0} \int_{1/(1+z)}^1 
   \frac{1}{\sqrt{\Omega_m a^{-3}+\Omega_\Lambda}}\frac{da}{a^2} \,.
\end{equation}
In this case the dipole model~(\ref{eq:dipolemonopole}) is modified to use \mbox{$B(r) = B_0\,d(z)$}. This parameterisation makes physical sense if one imagines that any spatial $\alpha$-variation is ``fixed'' to the CMB frame and follows the same scale factor, $a(t)$.

This paper is organised as follows. In Section~\ref{sec:AICc} we develop criteria to assess the relative likelihood of different models of $\alpha$-variation which we use throughout this work. After testing the normalcy of errors using a quantile-quantile plot, we use the modified Akaike Information Criterion to confirm that the preferred model of cosmological $\alpha$-variation is a dipole model with a monopole term. In Section~\ref{sec:dipole_significance} we perform some additional statistical tests to determine the significance of the dipole: an $F$-test and a modified error-ellipsoid method, described in detail in Appendix~\ref{sec:EEM}. To test the robustness of the data we perform biased and unbiased clippings of the data. In Section~\ref{sec:dipole_robustness} it is shown that \emph{at least} 40\% of the quasar absorption systems would need to be removed from the sample to reduce the significance of the dipole to $1\sigma$. 

The monopole term of our preferred model of $\alpha$-variation is somewhat difficult to reconcile with the requirement that $\delta\alpha/\alpha$ should be zero locally, therefore in Section~\ref{sec:magnesium} we test a hypothesis that changes in the relative isotope abundances of magnesium could lead to the observed monopole term. We find that a relatively small increase in the abundance of $^{26}$Mg relative to $^{24}$Mg in the absorbers could account for the monopole term. Finally, in Section~\ref{sec:wall} we test whether the data supports a different model: that a ``domain wall'' separates regions of the Universe with different values of $\alpha$ \citep[as proposed by][]{olive11prd}. Finding a best fit for this model requires a genetic algorithm, presented in Appendix~\ref{sec:genetic}.

\section{Astronomical data}
\label{sec:data}

The data used in this paper are taken from~\citet{murphy04lnp}~(Keck data) and~\citet{king12mnras}~(VLT), kindly supplied by the J. A. King in a usable text format including the location of each absorber (redshift and direction) and its measured value of $\Delta\alpha/\alpha$. In total there are 293 points in our data file including 140 from Keck and 153 from VLT of which seven absorbers are seen in both the Keck and VLT samples. The data is described fully in~\citet{king12mnras}, here we just note some points of particular interest to the analysis presented in this paper. Both samples account for unknown sources of scatter in the data by including extra systematic errors, $\sigma_\textrm{rand}$, that are added in quadrature with the underlying statistical error. We will take each sample in turn.

The Keck data is taken from the `fiducial sample' of~\citet{murphy04lnp}. This includes several samples of Keck/HIRES spectra observed independently by different groups \citep[see][]{murphy03mnras,murphy04lnp} for details). The combined data is divided into high-$z$ ($z>1.8$) and low-$z$ subsamples, and the high-$z$ subsample is further divided into `low-contrast' and `high-contrast', defined as systems where both strong and weak lines are fitted. The 27 systems that constitute the high-contrast sample show large scatter in $\Delta\alpha/\alpha$ values, which is attributed to weak components not being fitted when they're near the high optical depth edges of the strong transitions' profile. To this high-contrast sample is added an additional error $\sigma_\textrm{rand} = 1.743\times 10^{-5}$ which serves to reduce $\chi^2$ per degree of freedom $\chi^2/\nu$ to unity. The low-$z$ and low-contrast high-$z$ Keck samples do not require additional~$\sigma_\textrm{rand}$.

The VLT/UVES data presented in~\citet{king12mnras} also shows too much scatter in $\Delta\alpha/\alpha$. Quantifying this scatter depends on the underlying model assumptions, for example if one assumes that the dipole model is correct (i.e. that there is a physical dipole in $\alpha$ that is being observed) and the data supports this, then $\sigma_\textrm{rand}$ should be smaller than it would be for a monopole model. The data we use includes $\sigma_\textrm{rand} = 0.905\times 10^{-5}$. This is a conservative (large) value which will tend to reduce the significance of the dipole relative to a monopole model. In practice all models give values of $\sigma_\textrm{rand}$ in the same vicinity~\citep[see Table~2. of][]{king12mnras} and the results of this paper are relatively insensitive to the exact $\sigma_\textrm{rand}$ used.

\section{Information criteria for different models}
\label{sec:AICc}

In this work we use $\chi^2$ as a likelihood measure for the various models. For measured values $x_i = \Delta\alpha/\alpha$ and a model which takes values $\bar x_i=\bar x_i(\vec{r})$ at the $i$th quasar, this is defined as the sum of squared residuals $\xi_i$:
\begin{equation}
\label{eq:chisq}
\chi^2 = \sum_i \xi_i^2 = \sum_i \frac{(x_i - \bar x_i)^2}{\sigma_i^2} \,,
\end{equation}
where $\sigma_i$ is the observational uncertainty. $\chi^2$ corresponds to the negative of the log-likelihood function for the normally-distributed random variable $\xi$; smaller values of $\chi^2$ correspond to better model fits.

The normalcy of the residuals is an implicit assumption in using a $\chi^2$ statistic. The validity of such an assumption may be ascertained by way of a quantile-quantile, or Q-Q, plot, which compares the residuals of a data set against the quantiles of a given distribution, or another sample. In addition to determining whether a data sample is consistent with a given distribution, a Q-Q plot is a useful graphical method of comparing the qualitative features, such as skew, of samples of data against statistical distributions.

\begin{figure}[b]
  \centering
  \includegraphics[width=80mm]{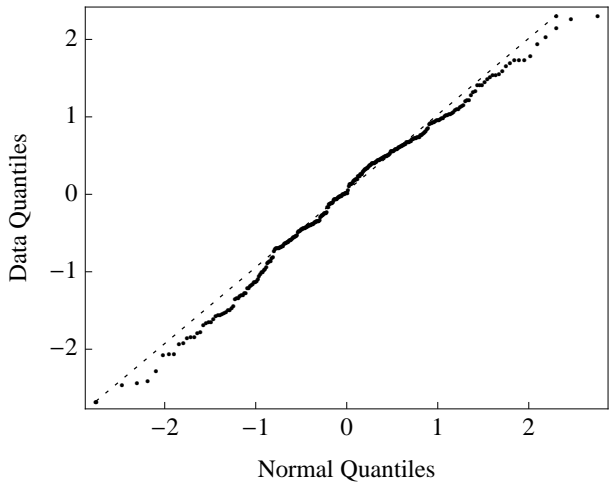}
  \caption{\label{fig:qqplot} Quantile-quantile plot for all $N=293$ data points. Coordinate $y_i$: ordered residuals, $\xi_i$, of the data. Coordinate $x_i$: empirical estimate of the location of the $i^{th}$ $N$-tile of the normal distribution, i.e.~$\textrm{CDF}(x_i)\approx i/N$ where $\textrm{CDF}(x)$ is the cumulative distribution function of the normal distribution. }
\end{figure}

We compare the residuals from the best fit dipole direction (light-travel distance) against a normal distribution in Figure~\ref{fig:qqplot}. If our residuals are normal, then we expect the plot to lie approximately across the diagonal. Our comparison, Figure~\ref{fig:qqplot}, suggests that the residuals are near-consistent with a normal distribution, though it should be observed that there is some `arching' in the curve, with the curve being below the diagonal near the tails and above near the centre. This indicates a negative skewness, that is, a long left-hand tail, in the residual distribution. However, in light of the relatively small size of our sample, the slight skew we observe does not necessarily indicate that the residuals are not normal. We conclude that $\chi^2$ is a reasonable statistic for our data.

The addition of free model parameters on any data set can only reduce $\chi^2$, therefore one must seek further justification for their inclusion. The modified Akaike Information Criterion ($\textrm{AIC}_\textrm{c}$; the modification is that it has been corrected for small sample sizes) penalises model parameters, allowing us to compare the goodness of a $\chi^2$ fit for different models. The $\textrm{AIC}_\textrm{c}$ is defined as~(\citealt{sugiura78cstm}; see also \citealt{liddle07mnras})
\begin{equation}
\label{eq:aicc}
\textrm{AIC}_\textrm{c} = \chi^2 + 2p + \frac{2p(p+1)}{n-p-1}
\end{equation}
where $p$ is the number of free parameters in our model and $n$ is the number of data points in the sample. The preferred model is the one with the lower $\textrm{AIC}_\textrm{c}$. A suggested interpretation scale for the $\textrm{AIC}_\textrm{c}$ is that introduced by \citet{jeffreys61book}, where a relative difference in the $\textrm{AIC}_\textrm{c}$ of two models of $<5$ indicates that the two models are almost equivalent in terms of significance.

In Table~\ref{tab:aicc} we present the results of our $\textrm{AIC}_\textrm{c}$ goodness-of-fit test of various $\alpha$-variation models to the data. On the basis of these results we find: the Keck sample exhibits a preference for the monopole fit; the VLT data is better fit by a dipole model; and the combined sample gives lowest $\chi^2$ for the dipole + monopole model and the two-value model. In the next section we present results from different statistical tests that suggest that the dipole model is preferred even for the Keck data subset.

\begin{table}[tb]
\caption{\label{tab:aicc} Goodness of fit test using $\chi^2/\nu$ ($\nu = n-p$ is the number of degrees of freedom) and the corrected Akaike Information Criterion ($\textrm{AIC}_\textrm{c}$) for various models of the form~(\ref{eq:dipolemonopole}). Models with a dipole term are calculated using both light-travel distance, $r$, and comoving distance, $d$. The ``two-value'' model allows separate monopole values for Keck and VLT. }
\begin{ruledtabular}
\begin{tabular}{lcdld}
Model  & $p$ & \multicolumn{1}{c}{$\chi^2$} & \multicolumn{1}{c}{$\chi^2/\nu$} & \multicolumn{1}{c}{$\textrm{AIC}_\textrm{c}$} \\
\hline
\multicolumn{5}{l}{\emph{Combined sample:}} \\
Null ($\Delta\alpha/\alpha = 0$) & 0 & 310.439 & 1.0595 & 310.439 \\
Monopole & 1 & 303.77 & 1.0403 & 305.784 \\
Dipole $r$ & 3 & 284.628 & 0.9815 & 290.711 \\
Dipole $d$ & 3 & 285.185 & 0.9834 & 291.268 \\
Dipole $r$ + monopole & 4 & 279.66 & 0.9677 & 287.799 \\
Dipole $d$ + monopole & 4 & 279.845 & 0.9683 & 287.984 \\
Two-value & 2 & 282.585 & 0.9711 & 286.626 \\
\hline
\emph{Keck (140 data points):}  & & & & \\     
Null & 0 & 157.898 & 1.1278 & 157.898 \\
Monopole & 1 & 132.799 & 0.9554 & 134.828 \\
Dipole $r$ & 3 & 142.439 & 1.0397 & 148.615 \\
Dipole $d$ & 3 & 143.611 & 1.0482 & 149.787 \\
Dipole $r$ + monopole & 4 & 131.28 & 0.9653 & 139.576 \\
Dipole $d$ + monopole & 4 & 131.309 & 0.9655 & 139.605 \\
\hline
\emph{VLT (153 data points):} & & & & \\       
Null & 0 & 152.541 & 0.9970 & 152.541 \\
Monopole & 1 & 149.786 & 0.9854 & 151.812 \\
Dipole $r$ & 3 & 139.908 & 0.9327 & 146.069 \\
Dipole $d$ & 3 & 138.839 & 0.9256 & 145.000 \\
Dipole $r$ + monopole & 4 & 139.288 & 0.9348 & 147.558 \\
Dipole $d$ + monopole & 4 & 138.167 & 0.9273 & 146.437 \\
\end{tabular}
\end{ruledtabular}
\end{table}

For the full (combined) sample, the best $\textrm{AIC}_\textrm{c}$ result (though not the best $\chi^2$) comes from the two-value model when $A_\textrm{Keck}=(-0.572\pm0.114)\times 10^{-5}$ and $A_\textrm{VLT}=(0.208\pm0.125)\times 10^{-5}$. According to the Jeffreys criterion, it is therefore of similar significance to the dipole+monopole model. The two-value model is expected if there exist unknown systematics that cause apparent shifts in $\alpha$, and these `intra-telescope' systematics are different for the two telescopes. However we will argue that the two-value result is not in conflict with the dipole result; it is actually expected given the distribution of quasars seen with Keck and VLT.

Let us first discuss the idea that there are unknown intra-telescope systematics that cause spurious shifts in the observed value of $\Delta\alpha/\alpha$. What kind of systematic is required? Any linear transform (offset or scaling) of the frequency scale will not mimic $\alpha$-variation. The many-multiplet method uses many lines for the analysis: some are positive shifters, some are negative shifters, and some are anchors. In order to mimic $\alpha$-variation it is therefore necessary for a systematic to shift lines in different directions. One suggestion is that intra-order systematic shifts in calibration of the Keck spectrograph could mimic $\alpha$-variation in any one absorber~\citep{griest10apj}; it is less clear that this would have a non-zero mean when averaged over many absorbers. Similar intra-order shifts have also been found in the VLT~\citep{whitmore10apj}, but it's worth noting that these are smaller in amplitude than the Keck shifts.

The proposition that intra-telescope systematics can mimic $\alpha$-variation can also be tested using the subset of quasar absorption systems that have data from both VLT and Keck telescopes. There are seven such systems, and a sophisticated analysis of them shows no systematic velocity offsets between the two telescopes~\citep{webb11prl}. Furthermore, the dipoles fit separately from the two telescopes point in the same direction (within $1\sigma$ errors). This is entirely unexpected if systematics accounted for monopole offsets for the telescopes: the probability that the dipoles would point in the same direction by chance was estimated at 6\%. A similar coincidence in the dipole directions was found when the sample was split into high ($z>1.6$) and low redshifts~\citep{webb11prl}; these subsamples use very different transitions in different ions with different responses to $\Delta\alpha/\alpha$~\citep[see, e.g.][]{murphy03mnras}.

\begin{figure}[tb]
  \centering
  \caption{\label{fig:dipolehistogram} A histogram of the distances, $r=ct$, to each absorber projected along the dipolar axis. The bars with positive-sloped lines on the left are from the Keck sample, while the bars with negative-sloped lines on the right represent the VLT data. It is clear that the two telescopes sample different portions of the sky.}
    \includegraphics[width=80mm]{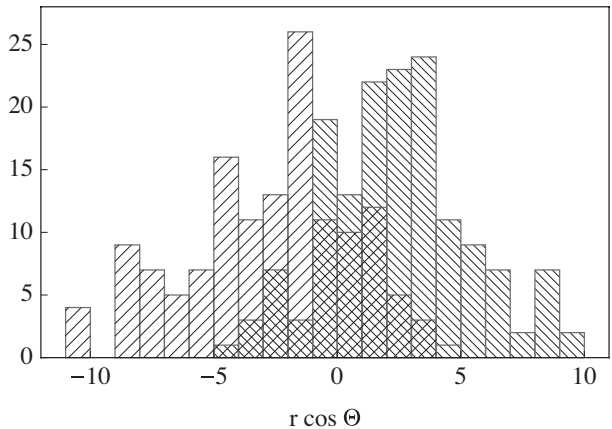}
\end{figure}

An alternative explanation to intra-telescope systematics is that the two-value result is simply an artefact due to the two telescopes sampling data from significantly different portions of the sky. If each telescope had full coverage of the sky, then one would expect that averaging over all absorbers would give back the monopole term of $\Delta\alpha/\alpha$. However, the telescopes do not each have full sky coverage. Figure~\ref{fig:dipolehistogram} shows the number of quasar absorption systems seen by each telescope against the distance to each absorber projected along the dipolar axis. The two telescopes gather data from different parts of the sky (this is simply due to their locations on the Earth). If the fitted dipole+monopole model represents the true distribution of $\alpha$ in the Universe, then it is to be expected that a weighted mean over the observed values yields different values for the two telescopes.

\section{Dipole significance}
\label{sec:dipole_significance}

Additional statistical tests are available for distinguishing between alternate ``nested'' models. Two models with $p_1$ and $p_2$ parameters ($p_1 \leq p_2$) are nested if the entire parameter space of model~1 is contained within that of model~2; this is obviously the case for the 4-parameter dipole+monopole model and the single-parameter monopole model. $\chi^2$ must improve when the three additional dipole parameters are included. The $F$-test provides us a method to find the probability that this improvement is due to chance. We define the $F$ statistic by
\begin{equation}
F = \frac{(\chi^2_1 - \chi^2_2)/(p_2 - p_1)}{\chi^2_2/(n - p_2)}\,,\quad p_2 > p_1
\end{equation}
where $n$ is the number of data points. Under the null hypothesis this statistic has an $F$-distribution; we can infer the probability that the improvement in $\chi^2$ is due to chance by
\begin{equation}
P = 1 - \int_0^F F(x;\, p_2 - p_1,\, n - p_2)\,dx \,.
\end{equation}
We express this as a standard deviation $\sigma$ in the usual way, $1-P = \erf(\sigma/\sqrt{2})$ where $\erf$ is the Gauss error function. The significance of the dipole for the Keck, VLT, and combined data samples is presented in Table~\ref{tab:sigRes}. We again find that the Keck sample shows no preference for a dipole model, the VLT sample shows a significant preference for a dipole, and the combined sample shows a strong preference for a dipole model.

\begin{table}[tb]
\caption{\label{tab:sigRes} Significance of the dipole+monopole model using lookback time $r$ as the distance measure for the Keck sample, the VLT sample, and the combined data. Significance is obtained using both the $F$-test and the error-ellipsoid method (\textrm{EEM}).}
\begin{tabular}{cccc}
\hline\hline
 & Keck & VLT  & Combined \\
\hline
$F$-test ($\sigma$) & 0.43 & 2.50 & 4.21 \\
\textrm{EEM}\ ($\sigma$)  & 0.55 & 2.37 & 4.09 \\
\hline
\end{tabular}
\end{table}

Another method of assessing the significance of the dipole model is to simply calculate the usual statistical significance of the dipole parameters relative to zero. We use a maximum likelihood analysis to find the preferred direction and magnitude of the dipole. An error ellipsoid is generated representing the uncertainty in the three-dimensional parameter space of the dipole, and the proportion of the best-fit dipole that is contained within this error ellipsoid indicates the significance of the dipole model. This method was used, for example, by \citet{cooke10mnras} to assess evidence for a dipole in the accelerated expansion of the Universe. In Appendix~\ref{sec:EEM} we present an extension of this method, which we refer to as the error-ellipsoid method (\textrm{EEM}), to allow for distance-dependent dipole models of the type~(\ref{eq:dipolemonopole}). Table~\ref{tab:sigRes} shows that the \textrm{EEM} gives values for the significance of the dipole+monopole model relative to the monopole-only model which are consistent with the $F$-test method for all data sets. 

\section{Robustness of the dipole model}
\label{sec:dipole_robustness}

We wish to test whether the dipole result is due to a small proportion of the data sample. To do so we perform biased and unbiased clippings of the data. We start with a biased iterative clipping method, where at each increment we remove a point that lends support to the dipole model. One option is to remove the absorber producing the smallest residual value $\xi$. However, the data points with small residuals may be absorbers lying near the plane orthogonal to the fitted dipole axis, and these would not contribute to the dipole effect. Therefore we introduce an angular weighting that further biases the clipping towards removal of points that lie on the dipole axis:
\[
\frac{\xi_i}{\cos\theta_i} =
\frac{x_i - \bar{x_i}}{\sigma_i \cos \theta_i} \,,
\]
where the $\bar{x_i}$ is the dipole model value for that absorber and $\theta_i$ is the angle from the dipole axis.

\begin{figure}[tb]
  \centering
  \caption{\label{fig:smallclip} Significance of the preference for a dipole+monopole model over the monopole model versus number of data points clipped. At each step we have removed the absorber with smallest weighted residual $\xi/\cos\theta$, where $\theta$ is the angle between the location of the absorber and the dipole axis. Solid line: significance assessed using $F$-test; dashed line: significance assessed using \textrm{EEM}. The vertical dashed line indicates the point at which the dipole model is not preferred according to the $\textrm{AIC}_\textrm{c}$.}
  \includegraphics[width=80mm]{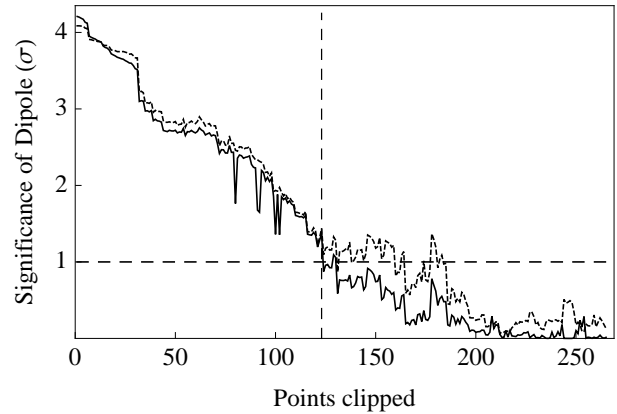}
\end{figure}

After each clipping of the absorber with smallest weighted residual, we calculate the significance of the dipole using the $\textrm{AIC}_\textrm{c}$, $F$-test and \textrm{EEM}. We present the result of these tests in Figure~\ref{fig:smallclip}. The solid black line gives the significance of preference for the dipole+monopole model over the pure monopole model as calculated via the F-test. This test indicates that the dipole model is preferred over the monopole (at $1\sigma$ significance or more) until approximately 120 of the strongest points, or 40$\%$ of the data, is removed. The significance as calculated using \textrm{EEM} is given by the dashed line of Figure~\ref{fig:smallclip}: we see that the consistency between significance assessed using $F$-tests and \textrm{EEM} established in Section~\ref{sec:dipole_significance} extends to clipped subsets of the data until $\sim120$ points have been removed and the significance is smaller than $1\sigma$. The vertical dashed line indicates the point at which the dipole+monopole model is no longer preferred over the monopole model as determined by the $\textrm{AIC}_\textrm{c}$. Again, this occurs after $\sim120$ points have been removed.

\begin{figure}[tb]
  \centering
  \caption{\label{fig:40pcremovals} Probability density function for the significance of the dipole after random removal of 40\% of data, assessed using the $F$-test. We see that in almost all cases the dipole+monopole model is preferred to the pure monopole model at more than $1\sigma$ significance.}
  \includegraphics[width=80mm]{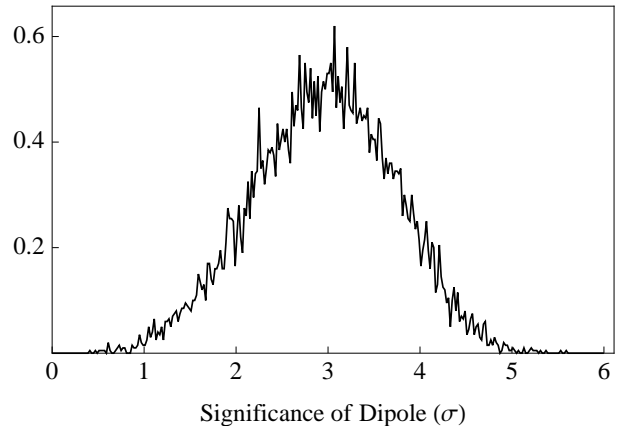}
\end{figure}

The results of our biased clipping suggest that approximately 40\% of the data that best supports the dipole must be removed before the significance of the dipole is reduced to $1\sigma$. Therefore, we expect that if we randomly remove 120 data points from the sample and re-fit our data, the dipole+monopole model would still be preferred (with at least $1\sigma$ significance) over the monopole model. In Figure~\ref{fig:40pcremovals} we present the probability density function for this unbiased clipping performed 10 000 times. Each time, the dipole significance is calculated using an F-test. We find that, as expected, the significance is at least $1\sigma$ in almost all 60\% subsets of the data. Taken together with our biased clipping, this shows that the dipole result is not due to a small number of outliers in the sample.

It is useful to contrast our biased iterative clipping to that presented in~\citet{webb11prl}. In that work, the point with \emph{largest} residual was removed at each step of the procedure. This is useful for showing that the dipole is not caused by a few outlying data points, but it can also remove absorbers that do not support the dipole model. For example, in~\citet{webb11prl} there is no weighting by $\cos\theta$, and deviant points that lie in the perpendicular plane may be removed. Therefore it is unsurprising that the significance of the dipole increases at first \citep[Fig. 4 in][]{webb11prl}. However, there is a trade-off: the significance eventually falls as more points are removed because there are simply fewer data. \citet{webb11prl} found that approximately 60\% of data with the largest residuals had to be discarded before the significance dropped below $3\sigma$.

\section{Magnesium and the Monopole}
\label{sec:magnesium}

We have seen from our $\textrm{AIC}_\textrm{c}$ analysis (Section~\ref{sec:AICc}) that there is some justification for including the monopole term of (\ref{eq:dipolemonopole}) as well as the dipole term. One natural interpretation of such a term is time-variation of the fine-structure constant, of the kind that was indicated in previous Keck studies~\citep{murphy03mnras}. One would then expect the monopole term to have an explicit time-dependence (or, equivalently, redshift-dependence $A = A(z)$), although the form may not be resolved by the data.

Another possible explanation for a monopole could be chemical evolution of the Universe, which changes the isotope abundance ratios. Since terrestrial abundances are assumed, any deviation from these ratios in the absorber will shift the centroid of the line profile, and this might mimic a change in $\alpha$. Accounting for a systematic such as this is difficult, especially since the isotope shift is unknown for many of the lines used in the analysis. The transition with the largest known shift used in the analysis are the $\lambda\lambda$2796 and 2803 lines in \ion{Mg}{ii}, and calculations suggest that the unknown isotope shifts are smaller and less important~\citep{kozlov04pra}.

We have applied a simple test to see whether the isotope abundance ratios of magnesium could cause the monopole. We have simply removed all data points taken from absorbers where the \ion{Mg}{ii} lines are present and used in the analysis. 113 absorbers remain from our initial sample of 293: our statistical significance is hugely reduced.  The removal of points with \ion{Mg}{ii} is by no means a random sampling: these \ion{Mg}{ii} lines are seen in low redshift systems since at $z \gtrsim 2$ they are redshifted outside the range of optical telescopes. 
 
The results of our fitting are shown in Table~\ref{tab:Mg_removed}. For models that include a dipole we use the light-travel time as a measure of distance $B(z) = B_0 r(z)$. We see that the best-fit dipole parameters for the \ion{Mg}{ii}-removed data are consistent with those of the complete data set, but we no longer have a statistically significant monopole. Using an $F$-test we find that the dipole+monopole model is preferred over the monopole model at $1.9\sigma$ significance for the \ion{Mg}{ii}-removed set.

\begin{table*}[tb]
\caption{\label{tab:Mg_removed} Best fit parameters and values of $\chi^2$ and $\chi^2/\nu$ for different model fits to the data when all absorbers with \ion{Mg}{ii} have been removed. In models that include a dipole, the direction of the dipole axis is specified using right ascension (RA) and declination (Decl.) in equatorial coordinates.}
\begin{ruledtabular}
\begin{tabular}{llldl}
\multicolumn{2}{c}{Model} & Parameter values & \multicolumn{1}{c}{$\chi^2$} & \multicolumn{1}{c}{$\chi^2/\nu$} \\
\hline
\multicolumn{5}{l}{\emph{Subset with \ion{Mg}{ii} data removed:}} \\
Null & $\Delta\alpha/\alpha = 0$ & --- & 126.565 & 1.1200 \\
Monopole & $\Delta\alpha/\alpha = A$ & $A = 0.16\,(15)\times 10^{-5}$ & 125.421 & 1.1198 \\
Dipole & $\Delta\alpha/\alpha = B_0\, r\cos\theta$
   & $B_0 = 0.105\,(35)\times 10^{-5}\,\text{Glyr}^{-1}$ & 117.491 & 1.0681 \\
  && $\text{RA} = 16.0\,(1.8)\ \textrm{hr}$, $\text{Decl.} = -63\,(12)^\circ$ \\
Dipole + monopole & $\Delta\alpha/\alpha = A + B_0\, r\cos\theta$
   & $A = 0.12\,(15)\times 10^{-5}$ & 116.863 & 1.0721\\
  && $B_0 = 0.103\,(35)\times 10^{-5}\,\textrm{Glyr}^{-1}$ \\
  && $\text{RA} = 15.8\,(1.8)\ \textrm{hr}$, $\text{Decl.} = -61\,(12)^\circ$ \\
\hline
\multicolumn{5}{l}{\emph{All data:}} \\
Dipole + monopole & $\Delta\alpha/\alpha = A + B_0\, r\cos\theta$
   & $A = -0.19\,(8)\times 10^{-5}$ & 279.66 & 0.9677 \\
  && $B_0 = 0.106\,(22)\times 10^{-5}\,\textrm{Glyr}^{-1}$ \\
  && $\text{RA} = 17.4\,(1.0)\ \textrm{hr}$, $\text{Decl.} = -62\,(10)^\circ$ \\
\end{tabular}
\end{ruledtabular}
\end{table*}

We may try to estimate the magnitude of the change in \ion{Mg}{ii} abundances that could mimic the observed monopole term in the original data set. The two \ion{Mg}{ii} lines (at $\sim 35000\,\textrm{cm}^{-1}$) are often observed simultaneously with \ion{Fe}{ii} lines ($\sim 40000\,\textrm{cm}^{-1}$) which in these systems will provide most of the sensitivity to $\alpha$-variation ($q \sim 1500$ for positive shifters in \ion{Fe}{ii}). To obtain our order-of-magnitude estimate we assume here that only \ion{Mg}{ii} and positive-shifting \ion{Fe}{ii} lines are present in the system.

The procedure of~\citet{webb11prl} is to simultaneously fit the redshift $z$ and $\Delta\alpha/\alpha$ (along with column densities, Doppler widths, etc.) for the entire quasar spectrum. The monopole term in the full data set, $\Delta\alpha/\alpha = -0.19\times 10^{-5}$, corresponds to a shift in the \ion{Fe}{ii} lines of
\[
\frac{\Delta\omega}{\omega}\Big|_{\ion{Fe}{ii}} = \frac{2q}{\omega} \frac{\Delta\alpha}{\alpha}
	\approx -1.4\times 10^{-7}
\]
This shift manifests as an effective change in the redshift of the quasar absorption system as measured by the \ion{Fe}{ii} lines, but not as measured by the \ion{Mg}{ii} lines. It is possible to obtain consistency in measured $z$ if the relative isotope abundances of \ion{Mg}{ii} are assumed to vary. The isotope shift of the \ion{Mg}{ii} lines is $\Delta\omega_{IS} = \omega^{26} - \omega^{24} = 0.102\,\textrm{cm}^{-1}$~\citep{drullinger80ap}. In order to compensate the monopole term of the observed $\alpha$-variation, the relative abundance of $x = \element{26}{Mg}/\element{24}{Mg}$ would have to change by
\begin{gather}
	x \frac{\Delta\omega_{IS}}{\omega}\Big|_{\ion{Mg}{ii}} = 1.4\times 10^{-7} \nonumber \\
	x \approx 0.05 \nonumber
\end{gather}
That is, a roughly 5\% increase in the relative abundance of \element{26}{Mg} could remove the observed monopole in $\Delta\alpha/\alpha$. Alternatively, a $\sim10\%$ increase in the \element{25}{Mg} abundance relative to \element{24}{Mg}, would also work -- or any combination of the two (e.g. the absorber may have the terrestrial $\element{25}{Mg}:\element{26}{Mg}$ ratio, and a reduction in \element{24}{Mg} relative to the heavier \element{25,26}{Mg} isotopes).

Of course our assumptions are quite rough. However a previous study of the effect of isotope abundance in the Keck data came to much the same conclusion~\citep{murphy04lnp}. In their method, the quasar absorption spectra were refitted using a different value of magnesium heavy-isotope abundance and the $\Delta\alpha/\alpha$ extracted were averaged assuming a monopole model. This found a linear relationship between their assumed $\element{25,26}{Mg}/\element{24}{Mg}$ ratio and the extracted $\Delta\alpha/\alpha$. A change in $\Delta\alpha/\alpha$ of $0.2\times 10^{-5}$ required an increase in the abundance of heavy Mg isotopes relative to \element{24}{Mg} of around 10\%~\citep[see Fig.~6 of][]{murphy04lnp}. A similar analysis applied to the $z < 1.6$ subsample of the Keck+VLT data also found that the monopole could be removed by a change in the $\element{25,26}{Mg}/\element{24}{Mg}$ abundance ratio from the terrestrial value of 0.21 to a heavy-isotope enhanced value of $0.32\pm0.03$~\citep{king12mnras}.

Generally other species are present in our quasar absorption spectra, and this could remove the degeneracy between variation of magnesium isotope abundance and variation of $\alpha$. A more complete analysis should allow Mg isotope abundances to vary in the fitting procedure, which may be possible if this is restricted to only one additional parameter in the entire sample of $\sim 300$ absorption systems.

\section{Domain wall model}
\label{sec:wall}

Recently a different model for spatial variation of $\alpha$ has been proposed by \citet{olive11prd} where the Universe is divided into two domains, each with a different value of $\alpha$. On the Earth-side of the domain wall $\alpha$ takes the terrestrial value, while on the other side it takes a different value. We can parametrize this model by the equation
\begin{equation}
\label{eq:wall_model}
\frac{\Delta \alpha}{\alpha} = \begin{cases}
  A, & d\cos\theta - d_\textrm{wall} > 0 \\
  0  & \textrm{otherwise}
	\end{cases}
\end{equation}
where $d_\textrm{wall}$ is the shortest distance to the wall (we use comoving distance, Eq.~(\ref{eq:comoving_distance}), in this section) and $\theta$ is the angle between the direction of the shortest distance to the wall and a quasar. 

Again, values for $d_\textrm{wall}$ and the direction of the wall are chosen to minimise $\chi^2$, however in this case $\chi^2$ is discontinuous with respect to changes in the parameters: any given quasar must either be on this side of the wall (model value $\Delta\alpha/\alpha = 0$) or the other (model value $\Delta\alpha/\alpha = A$). This necessitates the use of a genetic algorithm to determine the minimum values of $\chi^2$; our algorithm is presented in Appendix~\ref{sec:genetic}.

\begin{figure}[tb]
  \centering
  \caption{\label{fig:bestfitwall} $\chi^2$ for the wall model as a function of comoving distance to the wall, $d_\textrm{wall}$. Solid line: in the direction of the best fit for the wall (the best fit occurs at $d_\textrm{wall}= 5.513$~Gly); dashed line: in the direction of the dipole axis (decreasing $\alpha$ direction). By comparison, $\chi^2$ for the spatial gradient (dipole) model is 279.66.}
  \includegraphics[width=80mm]{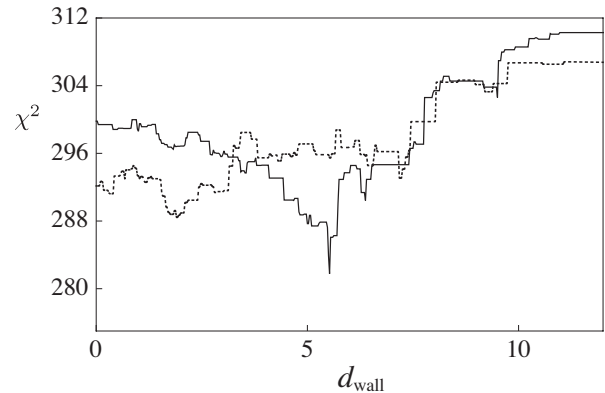}
\end{figure}

The best fit model for the entire data sample was found to have $A = -1.055\times 10^{-5}$, $d_\textrm{wall}= 5.513$~Gly, $\textrm{RA} = 20.1$~hr, $\textrm{Decl.} = 68^{\circ}$. The model has 4-parameters with $\chi^2 = 281.76$; this can be compared with the 4-parameter dipole+monopole model with comoving distance, for which $\chi^2 = 279.85$. However it is worth noting that this minimum $\chi^2$ is obtained for an extremely narrow range of parameters, and even a small deviation in any of them increases it dramatically. For example, in Figure~\ref{fig:bestfitwall} we present $\chi^2$ in the direction of best-fit (solid line) as a function of $d_\textrm{wall}$.

The best-fit wall is located at an angle $47^\circ$ to the dipole axis (towards negative values of $\Delta\alpha/\alpha$, i.e. $180^\circ$ from the directions presented in Table~\ref{tab:Mg_removed}). This is not consistent: if the Universe really does have a domain wall in values of $\alpha$ one would expect a dipole fit to the data to have the same direction. If we calculate $\chi^2$ as a function of wall distance parameter $d_\textrm{wall}$ along the direction that is given by the best-fit dipole we obtain the dashed line in Figure~\ref{fig:bestfitwall}. It is seen that the behaviour is much as would be expected if the dipole model is correct: as $d_\textrm{wall}$ increases, fewer absorbers are found on the other side of the wall, the model value $\Delta\alpha/\alpha=0$ for more absorbers, and $\chi^2$ increases.

\section{Conclusion}

We have developed a number of statistical tests to use quasar absorption data to differentiate between possible models of cosmological $\alpha$-variation. We confirm that a model of variation including both dipole and monopole terms is preferred, however the monopole may be an artefact due to non-terrestrial magnesium isotope ratios in the absorbers. A small change (approximately 5\%) in the ratio of \element{26}{Mg} to \element{24}{Mg} could account for the monopole, leaving the data to be explained by a purely spatial gradient in values of $\alpha$ across the Universe.

Our robustness tests on the data showed remarkable consistency between various statistical tests: at least 40\% of the data must be deleted in a heavily biased way to reduce the significance of the $\alpha$-dipole to $1\sigma$. An unbiased removal of 40\% of the data leaves the dipole significance at around $3\pm1\sigma$. This confirms that the original result, that there is evidence for a spatial gradient in values of $\alpha$ at the $4.1\sigma$ level, is not due to only a few deviant points.

Finally we developed a genetic algorithm to find best-fit parameters for the domain wall model of $\alpha$-variation suggested by~\citet{olive11prd}. We find that the best fit direction of the wall is not consistent with the dipole direction. This is because $\chi^2$ in the best-fit region is anomalously low with respect to its local neighbourhood of parameter space. On the other hand if we force the domain wall to be perpendicular to the best-fit dipole direction, the resulting $\chi^2$ are significantly larger than those of the dipole model.

\acknowledgments
We thank C. Angstmann, M.~T. Murphy, J.~A. King, J.~K. Webb, and F.~E. Koch for useful discussions. This work is supported by the Australian Research Council.

\appendix
\section{Assessing the dipole significance using error-ellipsoid method}
\label{sec:EEM}

In order to present a comparison of the significance of the models described above, we employ a similar strategy to that taken by~\citet{cooke10mnras}, generalised for our distance-dependent dipole models.
The probability that we observe data $x_i = (\Delta\alpha/\alpha)_i$ given our model value at the $i$th quasar, $\bar x_{i}$, is
\begin{equation}
\text{Pr}(x_i|\bar x_{i}) = \frac{1}{\sqrt{2\pi(\sigma^2+\sigma_i^2)}}~\exp\left( \frac{-(x_i-\bar x_{i})^2}{2(\sigma^2+\sigma_i^2)} \right) \,,
\end{equation}
where $\sigma_i$ is the observational error and $\sigma$ represents an intrinsic scatter of the true values from our model values $\bar x_i$. The best fit model parameters can then be found so that they maximise the log-likelihood function
\begin{align}
\mathcal{L} &= \ln \left[ \prod_i \text{Pr}(x_i |\bar x_{i}) \right] \nonumber \\
   &= \sum_i \ln \left( \frac{1}{\sqrt{2\pi(\sigma^2+\sigma_i^2)}} \right)
        - \frac{(x_i-\bar x_{i})^2}{2 (\sigma^2+\sigma_i^2)} \,.
\label{eq:loglikelihood}
\end{align}

For a dipole model, $\bar x_{i}$ is some function on the dipole direction and magnitude, which we represent as {\bf L}. Maximising the likelihood with respect to the model parameter gives the expression
\[
\bf A \cdot L = V. 
\]

The covariance matrix for $\bf L$ is given by ${\bf A}^{-1}$, which will depend on the various models as follows:
\begin{align*}
\bar x_{i} &= m + B\,\text{cos}\,\theta
  & \bar x_{i} &= m(1+\pmb{\hat{l}}_i\cdot \mathbf{L} ) \\
& & |{\bf L}|  &= B/m, \\
& & A          &= \sum_i \frac{m^2}{\sigma^2+\sigma^2_i} (\pmb{\hat{l}}_i \otimes \pmb{\hat{l}}_i).
\end{align*}
\begin{align*}
\bar x_{i} &= Br\,\text{cos}\,\theta
  & \bar x_{i} &= r \; \pmb{\hat{l}}_i \cdot \mathbf{L} \\
& & |{\bf L}|  &= B, \\
& & A          &= \sum_i \frac{r^2}{\sigma^2+\sigma^2_i} (\pmb{\hat{l}}_i \otimes \pmb{\hat{l}}_i).
\end{align*}
\begin{align*}
\bar x_{i} &= m + Br\,\text{cos}\,\theta
  & \bar x_{i} &= mr(1/r+ \pmb{\hat{l}}_i \cdot \mathbf{L}) \\
& & |{\bf L}|  &= B/m, \\
& & A          &= \sum_i \frac{(mr)^2}{\sigma^2+\sigma^2_i} (\pmb{\hat{l}}_i \otimes \pmb{\hat{l}}_i).
\end{align*}
\begin{align*}
\bar x_{i} &= B\,\text{cos}\,\theta
  & \bar x_{i} &= \pmb{\hat{l}}_i\cdot \mathbf{L} \\
& & |{\bf L}|  &= B, \\
& & A          &= \sum_i \frac{1}{\sigma^2+\sigma^2_i} (\pmb{\hat{l}}_i \otimes \pmb{\hat{l}}_i).
\end{align*}
where $m$ is the monopole magnitude, $B$ is the dipole magnitude, $r$ is the distance to the absorber and $\theta$ is the angle between direction of the dipole and the absorber. 

We have taken a direct computational approach to determining the best fit model parameters. The probability that a dipole is preferred over the monopole, and hence the significance of the dipole result, is then determined as with the method of Cooke and Lynden-Bell. Given the covariance matrix {\bf A$^{-1}$}, we have that the semiprincipal axes of the error ellipsoid for {\bf L} are $s_i = \sqrt{\lambda_i}\,{e}_i$, where $\lambda_i$ and ${e}_i$ are the eigenvalues and eigenvectors of the covariance matrix, respectively. The probability that the dipole should be included in the model, Pr({\bf L}), is then found by integrating over the volume of the parameter space obtained by expanding or contracting the error ellipsoid such that it just encloses {\bf L}. Then
\[
\text{Pr}({\bf L}) = \erf(1/\mu\sqrt{2}) - \sqrt{\frac{2}{\mu^2\pi}}\exp(-1/2\mu^2) \,,
\]
where $\mu = L_{err}/|\bf L|$ is the fraction of {\bf L} enclosed by the error ellipsoid
\[
L_{err} = [ {\bf (\hat{L}\cdot s_1)}^2 + {\bf (\hat{L}\cdot s_2)}^2 +{\bf(\hat{L}\cdot s_3)}^2  ] ^{1/2}
\]
and $\erf$ is the error function.

\section{Genetic algorithm for the wall model}
\label{sec:genetic}

The standard procedure of minimising $\chi^2$ to find best-fit model parameters is numerically difficult for the wall model~(\ref{eq:wall_model}) because $\chi^2$ is discontinuous with respect to the model parameters: $\chi^2$ changes in discrete steps depending on whether any given quasar absorption system is on the Earth-side or far-side of the wall. This necessitates a genetic algorithm with simulated annealing to perform the fit.

In a genetic algorithm, sets of model parameters are produced in `generations' with those producing the best fit (in our cases the smallest $\chi^2$) in a particular generation retained to `breed', producing the next brood of parameters. In each generation, `mutations' are introduced in order to produce the quasi-random variations of parameters necessary to cover the parameter space. The magnitude of such `mutations' are incrementally reduced in a process of simulated annealing so that the mutations become finer as parameters converge to their best fit values. As is usual when using genetic algorithms to determine best fits, several annealings are required. Unusually, in this case due to the discontinuous nature of the parameter space, we not only require several annealing steps but also several cycles of sets of annealing with new `isolated' populations of parameters in order to avoid a situation where the entire population inhabits local minima in the parameter space.

In each cycle, we perform 5 annealing steps, as it was found that subsequent steps produced limited exploration of the parameter space due to the rapid convergence of the parameters, and that the re-initialising of genetic material in each cycle produced more efficient coverage of the parameter space. For this reason, only a single survivor is retained per cycle, to be bred with nine new randomly created populations. In each annealing step, 500 generations of model parameters are produced, with the variance scaled down with each generation. 

We retain 10 sets of model parameters $\{i\}$ per generation, with which to produce 500 offspring. We can think of these as ``vectors'' in the 4-dimensional parameter space of the wall model. The offspring have genes taken to be the average of the parent values $\{i\}$ and $\{j\}$ with an additional random term:
\[
\{\textrm{child}\} = \frac{\{i\}+\{j\}}{2} + \{\xi\} \cdot \big(\{i\}-\{j\}\big)
\]
where the four terms in $\{\xi\}$ are drawn randomly from a Gaussian distribution with standard deviation $\sigma_\textrm{extra}$. The random scatter is reduced as the parameter values converge during the simulated annealing:
\[
\sigma_\textrm{extra} = \sigma_\textrm{max} \frac{500 - g}{500}
\]
where $g$ is the number of generations that have been run on a given annealing. The new generation of parameter values (parents and offspring) are then sorted by the fitness parameter $-\chi^2$ such that the 10 parameter sets producing the largest fitness parameters are retained.

After sufficient cycles (5 was found to be adequate) the result is tested for convergence. The result is deemed to be converged if the algorithm has not produced a better result for four consecutive cycles. Otherwise, the cycles continue. This sacrifices speed of convergence for robustness of the result. 

%\bibliographystyle{apsrmp4-1}
%\bibliography{references}

%merlin.mbs apsrmp4-1.bst 2010-07-25 4.21a (PWD, AO, DPC) hacked
%Control: author (75) reversed first initials jnrlst

\end{document}